\documentclass[preprint,showpacs,preprintnumbers,amsmath,amssymb,nofootinbib]{revtex4-1}
\usepackage[dvips]{graphicx}
\usepackage{graphicx}
\usepackage{amsfonts}
\usepackage{bm}
\usepackage{amsmath}
\usepackage{amssymb}
\usepackage{color}
\usepackage[all]{xy}

\def\be{\begin{equation}}
\def\ee{\end{equation}}
\def\bea{\begin{eqnarray}}
\def\eea{\end{eqnarray}}

\begin{document}
\title{Geometrothermodynamics of van der Waals systems }

\author{Hernando Quevedo}
\email{quevedo@nucleareas.unam.mx}
\affiliation{Instituto de Ciencias Nucleares, Universidad Nacional Aut\'onoma de M\'exico, AP 70543, Ciudad de M\'exico 04510, Mexico}
\affiliation{Dipartimento di Fisica and ICRA, Universit\`a di Roma ``Sapienza", I-00185, Roma, Italy}
\affiliation{Institute of Experimental and Theoretical Physics,
    Al-Farabi Kazakh National University, Almaty 050040, Kazakhstan}

\author{Mar\'\i a N. Quevedo}
\email{maria.quevedo@unimilitar.edu.co}
\affiliation{
    Departamento de Matem\'aticas,  Facultad de Ciencias B\'asicas,
    Universidad Militar Nueva Granada, Cra 11 No. 101-80,
    Bogot\'a D.C., Colombia}

\author{Alberto S\'anchez}
\email{asanchez@ciidet.edu.mx}
\affiliation{Departamento de Posgrado, CIIDET,\\
    {\it AP752}, Quer\'etaro, QRO 76000, MEXICO}

\begin{abstract}
We explore the properties of the equilibrium space of van der Waals thermodynamic systems. We use an invariant representation of the fundamental equation by using the law of corresponding states, which allows us to perform a general analysis for all possible van der Waals systems. The investigation of the equilibrium space is performed by using the Legendre invariant formalism of geometrothermodynamics, which guarantees the independence of the results from the choice of thermodynamic potential. We find all the curvature singularities of the equilibrium space that correspond to first and second order phase transitions. We compare our results with those obtained by using Hessian metrics for the equilibrium space. We conclude that the formalism of geometrothermodynamics allows us to determine the complete phase transition structure of systems with two thermodynamic degrees of freedom.   
\end{abstract}
\relax

\maketitle


\section{Introduction}
\label{sec:int}

Differential geometry is an important mathematical discipline that has found very broad applications in physics, chemistry, and engineering  \cite{frankel}. For instance, in Nature there exist so far only four fundamental interactions and all of them can be described by using concepts of differential geometry such as manifolds, metrics, connections, fiber bundles, etc.

In thermodynamics, differential geometry has been also applied intensively, providing an alternative way to describe thermodynamic systems. The starting point is the equilibrium space that is an abstract space whose points can be interpreted as representing equilibrium states of the system. One of the first attempts to introduce differential geometry in thermodynamics was proposed by Rao \cite{rao45} by introducing a Riemannian metric related to Fisher's information matrix. Rao's proposal has found many applications in statistical physics, information theory, and thermodynamics \cite{amari2012}. Furthermore, Hessian metrics can also be postulated to describe the geometric properties of the equilibrium space. In particular,
Weinhold \cite{wei75} and Ruppeiner  \cite{rup79,rup95} proposed to use the Hessian of the thermodynamic potentials known as internal energy and entropy, respectively, as metrics for the equilibrium space. 
The theoretical framework  in which Hessian metrics are applied to describe the equilibrium space is usually referred to as thermodynamic geometry.

More recently, one of us \cite{quev07} proposed the alternative approach of geometrothermodynamics (GTD),  in which the metrics of the equilibrium space should be independent of the choice of thermodynamic potential. This is a characteristic of classical equilibrium thermodynamics \cite{callen}, which implies that the properties of a system are invariant with respect to changes of the thermodynamic potential, i.e., invariant with respect to Legendre transformations. Another important characteristic of equilibrium thermodynamics is that all the properties of the systems are completely determined by the corresponding fundamental equations \cite{callen}. In GTD, the existence of a fundamental equation and Legendre invariance are the main assumptions that are used to construct the equilibrium space.

Thermodynamic geometry and GTD have been used to investigate ordinary laboratory systems as well as more exotic systems like black holes; see, for instance,  \cite{weinholdbook,rup14,abp15,manbis15,braluo14,hendi16,kubman15}.
In this work, we apply the formalism of GTD to investigate the physical properties of van der Waals systems. In particular, we clarify the relationship between the three different Legendre invariant metrics that are known in GTD. In fact, we find the curvature singularities of the corresponding equilibrium space and show that they correspond to the complete phase transition structure of van der Waals systems. We compare our results with those obtained by using Hessian metrics and show that whereas the GTD metrics contain information about the first and second order phase transitions, Hessian metrics are related to first order phase transitions only.

This work is organized as follows. In Sec. \ref{sec:gtd}, we present the basic ingredients of GTD. Section \ref{sec:rfeq} is dedicated to the derivation of the reduced fundamental equation of van der Waals systems. Then, in Sec. \ref{sec:gtdvdw}), we compute the GTD metrics and their curvature scalars to determine the locations in equilibrium space, where singularities exist. Furthermore, in Sec. \ref{sec:phases}, we show that the curvature singularities coincide with the points at which first and second order phase transitions take place. In Sec. \ref{sec:tdg}, we compute the Hessian metrics for van der Waals systems and show that they contain information about first order phase transitions. Finally, in Sec. \ref{sec:con}, we discuss our results.

\section{The formalism of geometrothermodynamics}
\label{sec:gtd}

In equilibrium thermodynamics, to describe a  system with $n$ thermodynamic degrees of freedom, we use $n$ extensive variables $E^a$ ($n=1,2,...,n$), $n$ intensive variables $I_a$, and a thermodynamic potential $\Phi$. Usually, the fundamental equation, from which all properties of the system can be derived, is given as a function $\Phi=\Phi(E^a)$ \cite{callen}. Then, the first law of thermodynamics can be expressed as (we assume the convention of summation over repeated indices)
\be
d\Phi =  I_a dE^a \ ,\quad I_a = \frac{\partial \Phi}{\partial E^a}
\label{flaw}
\ee
so that the intensive variables are functions of the extensive variables,  $I_a = I_a (E^a)$, representing the equations of state of the system. Moreover, it is assumed that the function $\Phi(E^a)$ behaves in accordance with the second law of thermodynamics.


In GTD, we use as starting point the $n$-dimensional equilibrium space ${\cal E}$ with coordinates $E^a$ so that each point of ${\cal E}$ can be interpreted as representing an equilibrium state of the system. To handle the Legendre invariance property of equilibrium thermodynamics as a geometric property of the GTD formalism, we represent Legendre transformations as coordinate transformations in an auxiliary $2n+1$ dimensional space called phase space ${\cal T}$ with coordinates $Z^A = \{\Phi, E^a, I_a \}$. Then, a Legendre transformation can be defined as  \cite{arnold,alberty1994}
\be
\{Z^A\}\longrightarrow \{\widetilde{Z}^A\}=\{\tilde \Phi, \tilde E ^a, \tilde I ^ a\}\ ,
\ee
\be
\Phi = \tilde \Phi - \delta_{kl} \tilde E ^k \tilde I ^l \ ,\quad
E^i = - \tilde I ^ {i}, \ \
E^j = \tilde E ^j,\quad
I^{i} = \tilde E ^ i , \ \
I^j = \tilde I ^j \ ,
\label{leg}
\ee
where $i\cup j$ is any disjoint decomposition of the set of indices $\{1,...,n\}$,
and $k,l= 1,...,i$. In particular, for $i=\{1,...,n\}$ and $i=\emptyset$, we obtain
the total Legendre transformation and the identity, respectively. We say that a geometric object is Legendre invariant if its functional form is not affected by a Legendre transformation. In particular, the contact 1-form $\Theta = d\Phi - I_a dE^a$, which according to Darboux theorem always exists in any odd dimensional manifold, is Legendre invariant because under a Legendre transformation it transforms as $\Theta\rightarrow \tilde\Theta = \tilde I_a d\tilde E^a$.
Furthermore, we equip the phase space with a Riemannian metric $G=G_{AB} dZ^A dZ^B$. Demanding Legendre invariance of the metric $G$, we obtain a set of algebraic equations for the components $G_{AB}$, whose solutions can be split into three different metrics \cite{qqs19}, namely
\be
G^{{I/II}} = (d\Phi - I_a d E^a)^2 + (\xi_{ab} E^a I^b) (\chi_{cd} dE^c dI^d) \ ,
\label{gupap}
\ee
which are invariant under total Legendre transformations.
Here $\xi_{ab}$ and $\chi_{ab}$ are diagonal constant $(n\times n)$-matrices.
For $\chi_{ab} = \delta_{ab}= {\rm diag}(1,\cdots,1)$, the resulting metric $G^{I}$ is used to describe systems with  first order phase transitions. Alternatively, for
$\chi_{ab} = \eta_{ab}= {\rm diag}(-1,\cdots,1)$, we obtain the metric $G^{{II}}$, which is used to describe  systems with second order phase transitions.  The third  Legendre invariant metric
\be
\label{GIII}
G^{{III}}  =(d\Phi - I_a d E^a)^2  + \sum_{a=1}^n E_a I_a   d E^a   d I^a \ ,
\ee
is invariant with respect to partial Legendre transformations\footnote{We notice that in previous works about the GTD metrics, we have considered a polynomial generalization of the metric (\ref{GIII}), in which the first term in the sum is $(E_aI_a)^{2k+1}$, where $k$ is an integer. However, to compare the explicit results of all the metrics $G^I$,  $G^{II}$, and $G^{III}$ for a particular thermodynamic system, it is necessary to set $k=0$ so that all the metrics are given in terms of quadratic polynomials in the variables $E_a$ and $I_a$}. 
Thus, the phase space in GTD is defined as the triplet $({\cal T}, \Theta, G)$, with  $\Theta$ and $G$ being Legendre invariant.

To guarantee that the equilibrium space ${\cal E}$ preserves the Legendre invariant property of the phase space, we assume that ${\cal E} \subset {\cal T}$ is defined  by the embedding map $\varphi: \cal E \rightarrow \cal T$ with $\varphi: \{E^a\} \mapsto \{\Phi(E^a), E^a, I^a(E^a)\}$ and $\varphi^*(\Theta)=0$. This last condition implies that on ${\cal E}$, the first law (\ref{flaw}) holds. Notice also that the definition of the embedding map $\varphi$ implies that the fundamental equation $\Phi=\Phi(E^a)$ must be known explicitly.
Moreover, it follows that  any metric $G$ in $\cal T$  induces a metric $g$ in $\cal E$ by means of the pullback $g=\varphi^*(G)$ or, in coordinates,
\be
g_{ab} = \frac{\partial Z^A}{\partial E^a} \frac{\partial Z^B}{\partial E^b} G_{AB}
\ .
\ee
Accordingly, in GTD there can be also three different metrics $g^{I}$, $g^{{II}}$, and $g^{{III}}$ for the equilibrium space, which can be represented as
\be
g^{{I/II}}_{ab} =   \beta_\Phi \Phi  \xi_a^{\ c}
\frac{\partial^2\Phi}{\partial E^b \partial E^c}   ,
\label{gdownf}
\ee
where $\xi_a^{\ c}=\delta_a^{\ c}={\rm diag}(1,\cdots,1)$ for $g^{I}$ and $\xi_a^{\ c}=\eta_a^{\ c}={\rm diag}(-1,1,\cdots,1)$ for
$g^{{II}}$. Moreover, the constant $\beta_\Phi$ represents the degree of homogeneity of the thermodynamic potential $\Phi$ \cite{qqs19}. The third metric of the equilibrium space can  be written as
\be
g^{{III}} = \sum_{a=1} ^n \left(\delta_{ad} E^d \frac{\partial\Phi}{\partial E^a}\right) \delta^{ab} \frac{\partial ^2 \Phi}{\partial E^b \partial E^c}
 dE^a dE^c \ .
  \label{gIII}
  \ee
As we can see from the above expressions, to find the explicit value of the metric $g$ of ${\cal E}$, we only need to know the fundamental equation $\Phi(E^a)$, which, as mentioned above, is part of the definition of the embedding map $\varphi$. This means that all the geometric properties of the equilibrium space can be derived from the fundamental equation. In this sense, GTD reproduces the property of equilibrium thermodynamics that all the information about the system can be derived from the fundamental equation. Following this idea, in the next section, we will explore the geometrothermodynamic properties of the equilibrium space in the case of  van der Waals systems.


\section{Reduced fundamental equation}
\label{sec:rfeq}

In this section, we will use the law of corresponding states to represent the fundamental equation in such a way that it can be applied to any van der Waals system. This  reduced and invariant representation of the fundamental equation allows us to derive all the thermodynamic properties of the system in a completely general manner \cite{greiner,huang,reif}.

The main van der Waals equation of state can be written as \cite{callen}
\bea \label{stateequation}
p=\frac{k_B
    T}{v-b}-\frac{a}{v^2}\,,
\eea
where $k_B$, $T$, and $v$ represent the Boltzmann constant, temperature, and volume per particle
$v=\frac{V}{N}$,  respectively. Using the critical point values $(p_c,v_c,T_c)$ as a scale, the above equation of state can be represented as
\be \label{cubic1} (v-v_c)^3=0\,,\ee
by choosing the van der Waals constants as
\be
a=3v_c{}^2p_c\,, \quad b = \frac{k_{{}_B} T_c}{8    p_c}\ .
\label{ab}
\ee
Furthermore, the compressibility factor $z_c = \frac{p_c v_c}{k_{{}_B}T_c}$ turns out to be a constant $z_c = \frac{3}{8}$, indicating the general applicability of the van der Waals equation of state, which is the essential feature of the law of corresponding states.

To derive the corresponding fundamental equation, we use the
partition function in the
canonical ensemble
\bea \label{partitionfunction}
Z(T,V,N)=\frac{1}{N!}\Bigg[\Bigg(\frac{2\pi m k_{{}_B} T }{h^2}
\Bigg)^{\frac{3}{2}}(V-bN)\exp{\left( \frac{a N}{V T}\right)}
\Bigg]^N\,,\eea
where $m$ is the mass of the particles and $h$ the Planck constant.
From here, we can calculate the  thermodynamic potentials corresponding  to
the free Helmholtz energy $F$, internal energy $U$ and entropy
$S$ as
\be
F=-k_{{}_B} T\ln{Z}\,,\ \ U=k_{{}_B} T^2\frac{\partial}{\partial T}\ln{Z}\,,
\ \ S=-\frac{\partial F}{\partial  T}\,.
\ee
Accordingly, we obtain the following expression for the entropy
\bea \label{entropy} s=k_{{}_B} \ln{(v-b)}+\frac{3}{2}k_{{}_B}
\ln{(u+\frac{a}{v})}+\frac{3}{2}k_{{}_B}
\ln{\left(\frac{4\pi m}{3h^2}\right)} + s_0 \,,
\eea
where
$u=\frac{U}{N}$ and $s=\frac{S}{N}$. Furthermore, substituting the
relationships (\ref{ab})  in the last  equation, we obtain
the reduced entropy
\bea \label{reducedentropy} \sigma=
\ln{\left(\omega-\frac{1}{3}\right)}+\frac{3}{2}
\ln{\left(\epsilon+\frac{3}{\omega}\right)}+\sigma_0 \,,
\eea
where we have defined the adimensional reduced
variables $\sigma=\frac{s}{k_{{}_B}}$, $\omega=\frac{v}{v_c}$,
and $ \epsilon=\frac{u}{u_c}$, in addition to the constant
$\sigma_0=\ln{(\frac{\pi m k_{{}_B} T_c}{2h^2})}$.
This is the reduced fundamental equation of a van der Waals system, in which we have used the law of corresponding states. The general and invariant character of this representation allows us to investigate together all the substances belonging to this class.

The fundamental equation must satisfy the first law of thermodynamics
\bea \label{firstlaw} du=Tds-pdv\,,
\eea
Introducing the reduced parameters in the form $u=u_c \epsilon$, $v=v_c \omega$, $T=T_c \tau$, $s=k_{{}_B}\sigma$, and $p=p_c\pi$ together with the
relationship the value of the compressibility
$z_c = \frac{3}{8} =  \frac{p_c v_c}{k_{{}_B}T_c}$
into (\ref{firstlaw}), we get the
reduced form of first law
\bea \label{reducefirstlaw2} d\sigma&=&\frac{3}{8\tau}d\epsilon
-\frac{3\pi}{8\tau}d\omega\,.\eea
Then, the  reduced  equilibrium conditions
\be
 \frac{3}{8\tau}=\Big( \frac{\partial
    \sigma}{\partial
    \epsilon}\Big)_\omega\,, \ \
\frac{3\pi}{8\tau}=\Big( \frac{\partial
    \sigma}{\partial \omega}\Big)_\epsilon\,,
\ee
lead to the expressions for the reduced temperature and pressure
\be
\tau=\frac{1}{4}\frac{(\epsilon \omega +3)}{\omega}\,, \ \
 \pi=8\tau\Big[\frac{1}{3\omega-1}+\frac{3}{2 \omega
    (\epsilon\omega+3)}\Big]\,\ee
from which it follows that
\bea \label{reducepressure} \pi=\frac{8
    \tau}{3\omega-1}-\frac{3}{\omega^2}\,.\eea
This is the reduced form of the main van der Waals equation of state
(\ref{stateequation}).


\section{Geometrothermodynamic properties}
\label{sec:gtdvdw}

As mentioned in Sec. \ref{sec:gtd}, in GTD there are three different metrics that are invariant with respect to Legendre transformations and can be used to describe the corresponding equilibrium space. In this section, we will investigate the main geometric properties that can be derived from each metric. In particular, we are interested in the behavior of the curvature as an indicator of the existence of phase transitions.

\begin{figure}[h]
\includegraphics[scale=0.3]{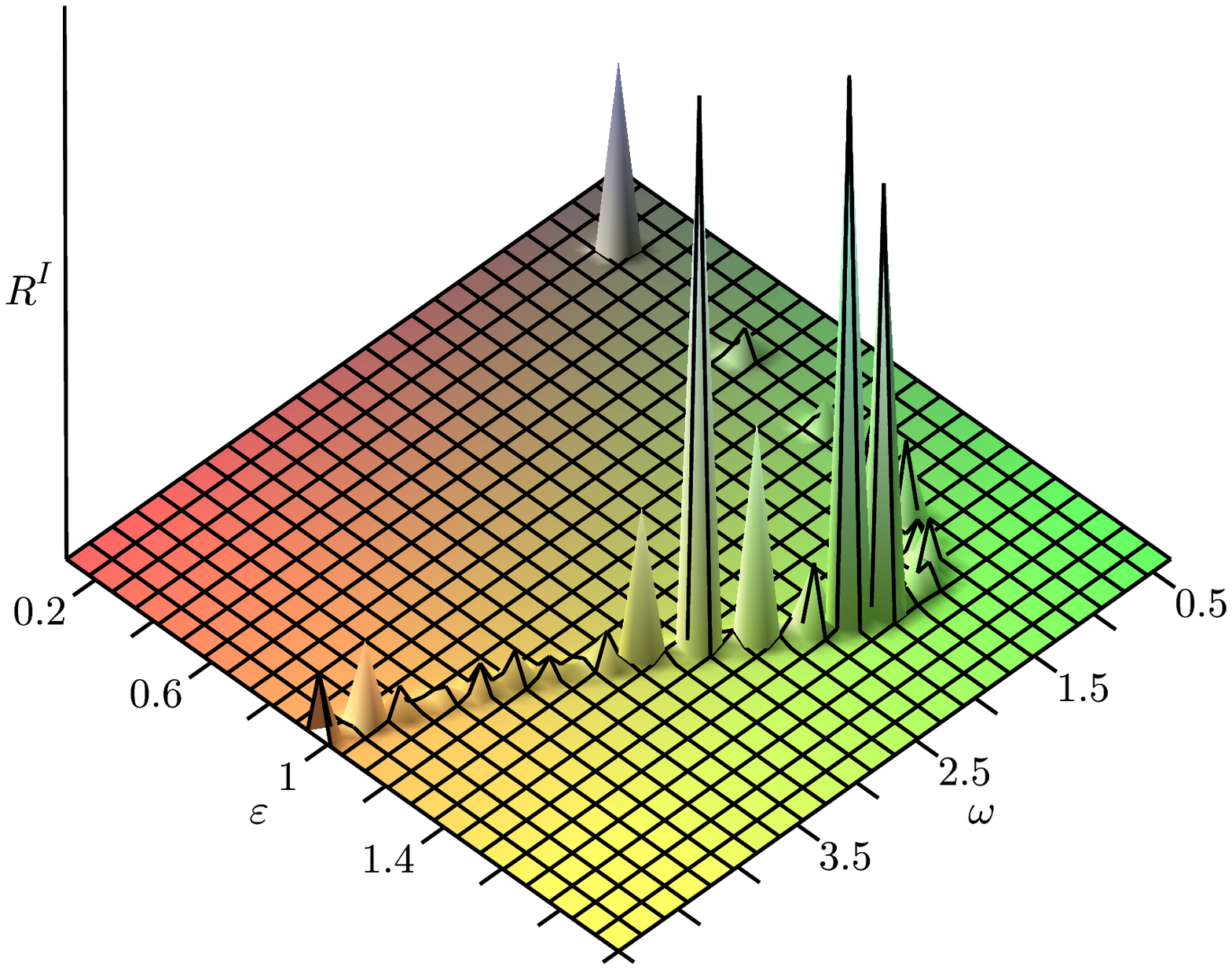}
\includegraphics[scale=0.3]{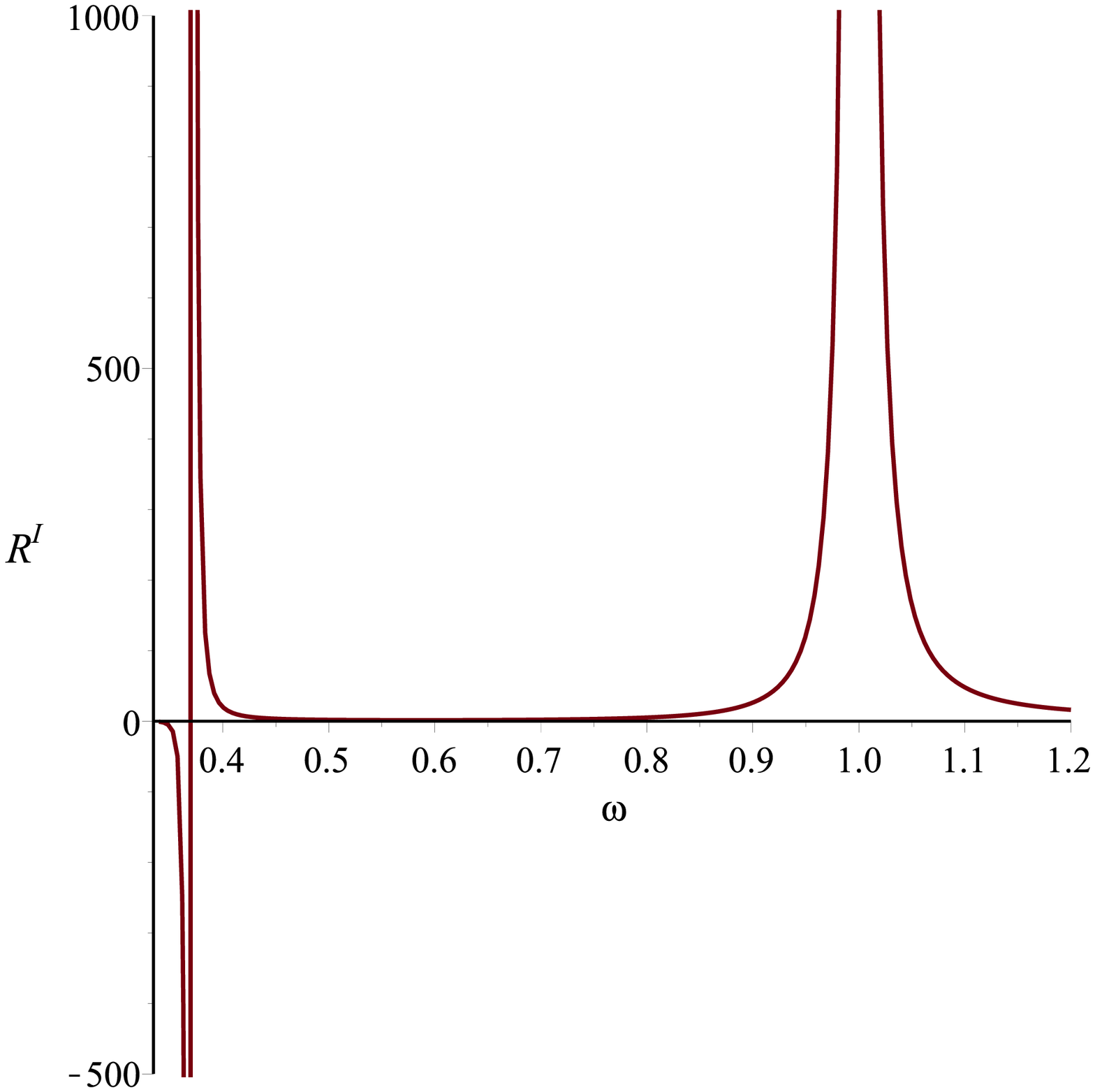}
\caption{Behavior of the scalar curvature of the metric $g^I$ as a function of the reduced energy $\epsilon$ and reduced volume $\omega$ (left panel) and as a function of the reduced volume $\omega$ with $\epsilon=1$. }
\label{fig1}
\end{figure}

Consider the metric $g^I$ for the equilibrium space
\be
g^I =    \Phi
\frac{\partial^2\Phi}{\partial E^b \partial E^c} \delta_a^{\ c}  dE^a dE ^b \ ,\ \  \delta_a^{\ c}  = {\rm diag}(1,1)\ ,
\label{gI}
\ee
where for simplicity we set in Eq.(\ref{gdownf}) the multiplicative constant $\beta_\Phi=1$, without loss of generality. In the case of the reduced van der Waals fundamental equation (\ref{reducedentropy}),
$\sigma=\sigma(\epsilon,\omega)$, we identify the thermodynamic variables as $\Phi=\sigma$ and $E^a=(\epsilon,\omega)$. Then,
\be
\label{reducedgtdIf}
g^{I}=\sigma\left(
\frac{\partial^2 \sigma}{\partial \epsilon^2}
d\epsilon^2+2\frac{\partial^2 \sigma}{\partial \epsilon
    \partial \omega } d\epsilon d\omega+\frac{\partial^2
    \sigma}{\partial \omega^2} d\omega^2\right)\,,
\ee
which for the reduced fundamental equation (\ref{reducedentropy}) becomes
\be
 \label{gtd1LCE}
 g^I =\frac{3
        \sigma}{2(\epsilon \omega+3)^2}\left[  -  \omega^2 d\epsilon^2
    +6 d\epsilon d\omega+3 f(\epsilon\,,\omega)d\omega^2\right]\,,
    \ee
where
    \bea \label{factoromega} f(\epsilon\,,\omega)=-\frac{2(\epsilon
        \omega+3)(\epsilon
        \omega^3-6\omega^2+6\omega-1)+3(3\omega-1)^2}{\omega^2
        (3\omega-1)^2}\,. \eea
The corresponding scalar curvature can be written as
    \bea \label{scalarI}
    R^I=\frac{(3\omega-1)^2(2\epsilon
        \omega^3-3\omega^2+6\omega-1)}{3\sigma (\epsilon
        \omega^3-6\omega^2+6\omega-1)^2}\,,\eea
from which we observe that curvature singularities occur for those values of $\epsilon$ and $\omega$ that satisfy the equation
\be
\epsilon
\omega^3-6\omega^2+6\omega-1 = 0.
\label{curvsingI}
\ee
In Fig. \ref{fig1}, we illustrate the behavior of the scalar curvature $R^I$. The picks on the graph correspond to singularities, i.e., locations where the curvature diverges and the conceptual basis of differential geometry breaks down.

\begin{figure}[h]
    \includegraphics[scale=0.3]{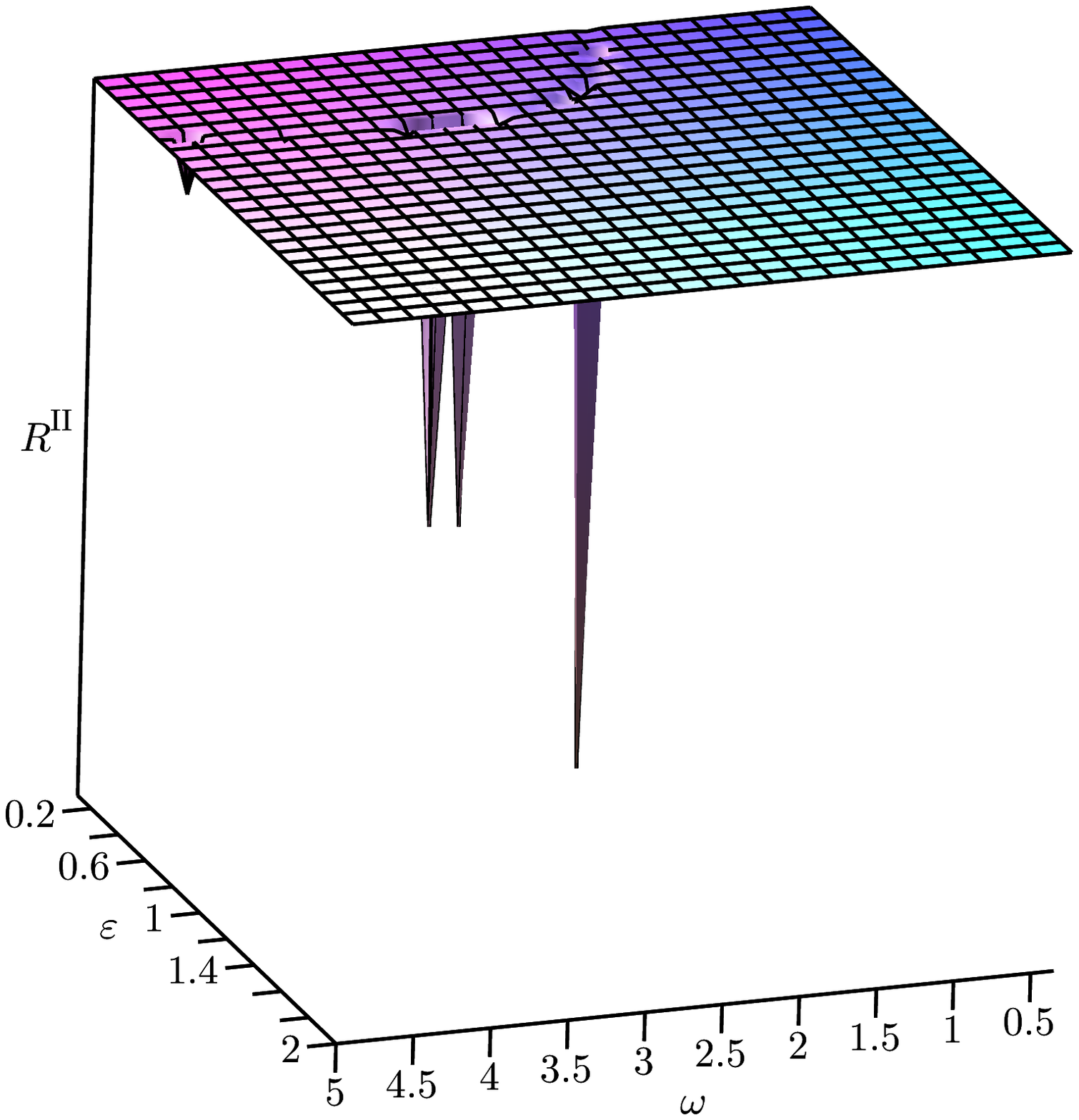}
    \includegraphics[scale=0.3]{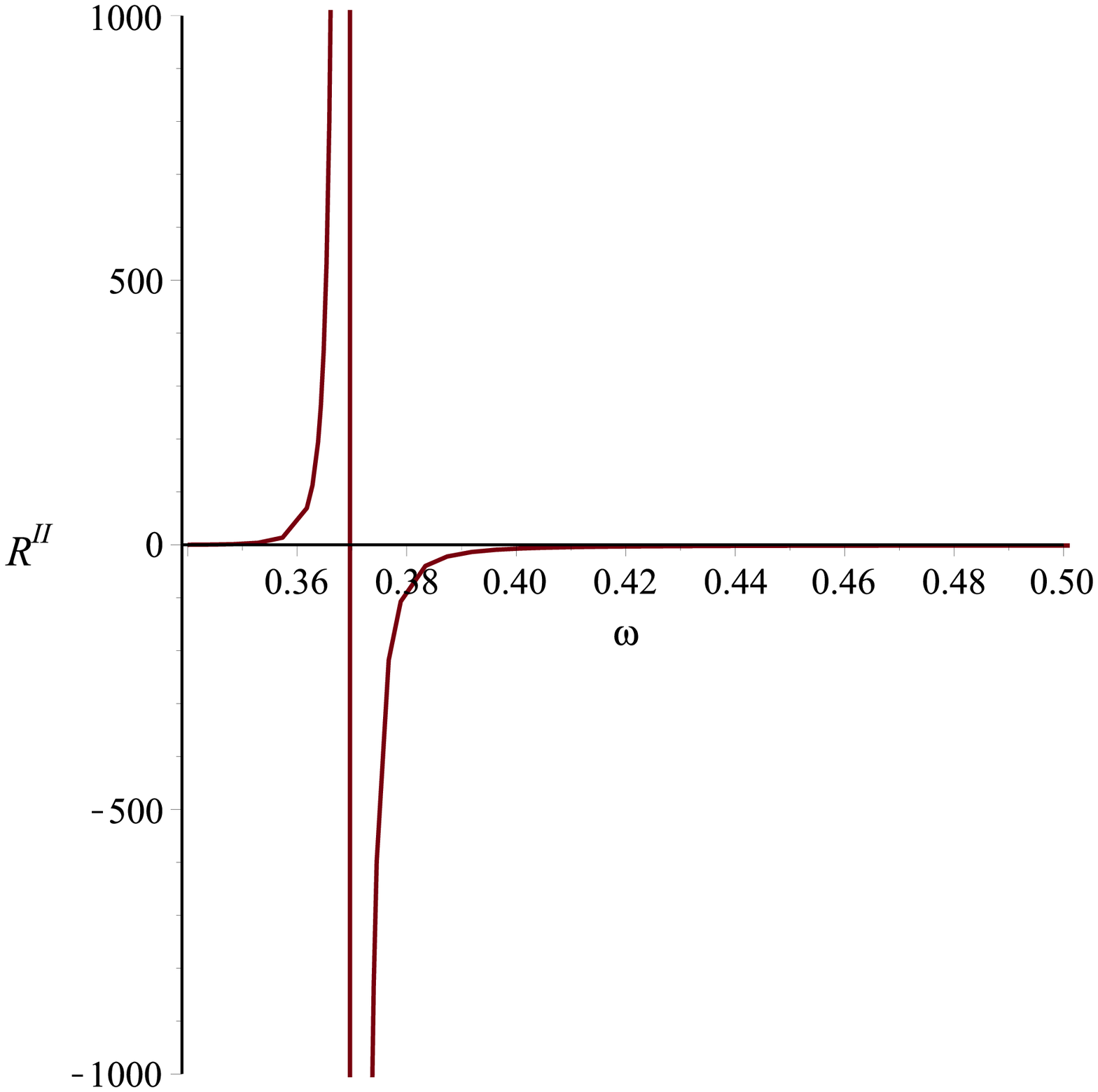}
    \caption{Behavior of the scalar curvature $R^{II}$ for several values of the variables $\epsilon$ and $\omega$ (left panel) and for $\epsilon=1$ and different values of $\omega$ (right panel). }
    \label{fig2}
\end{figure}

Consider now the GTD metric $g^{II}$ by assuming $\beta_\Phi = 1$ in Eq.(\ref{gdownf})
\be
g^{II} =   \Phi
\frac{\partial^2\Phi}{\partial E^b \partial E^c} \eta_a^{\ c}  dE^a dE ^b \ ,
\ \ \eta_a^{\ c} = {\rm diag}(-1,1)
\label{gII}
\ee
which for $\Phi=\sigma$ and $E^a=(\epsilon,\omega)$ reduces to
\bea \label{reducedgtdIIf} g^{II}=\sigma\left(-
\frac{\partial^2 \sigma}{\partial \epsilon^2}
d\epsilon^2+\frac{\partial^2 \sigma}{\partial \omega^2}
d\omega^2\right)\,.\eea
Furthermore, for the reduced van der Waals fundamental equation (\ref{reducedentropy}), we obtain
\bea
\label{gtd1LCEII}
g^{II}=\frac{3 \sigma}{2(\epsilon
    \omega+3)^2}\left[{\omega^2} d\epsilon^2 + 3
f(\epsilon\,,\omega)d\omega^2\right]\,,\eea
where the function $f(\epsilon,\omega)$ has been defined in Eq.(\ref{factoromega}). It is then straighforward to compute the scalar curvature, which can be written as
\bea \label{scalarII}
R^{II}= -\frac{8\omega^2
    (3\omega-1)(\epsilon \omega+3)^3}{3\sigma (2\epsilon^2 \omega^4
    -6\epsilon \omega^3 +12 \epsilon \omega^2-2\epsilon \omega
    -9\omega^2+18\omega-3)^2}\,,\eea
from which it follows that curvature singularities are determined by the equation
\be
2\epsilon^2 \omega^4
-6\epsilon \omega^3 +12 \epsilon \omega^2-2\epsilon \omega
-9\omega^2+18\omega-3 =0 \ .
\label{singII}
\ee
The behavior of the curvature scalar is depicted in Fig. \ref{fig2}. For comparison, we choose the same intervals of values for the variables $\epsilon$ and $\omega$. The singularity structure can easily be identified and is different from that of the scalar $R^I$.

Finally, we consider the metric $g^{III}$ given in Eq.(\ref{gIII}), which for $\Phi=\sigma$ and $E^a=(\epsilon,\omega)$ reduces to
\be
g^{III}= \epsilon \frac{\partial\sigma}{\partial \epsilon} \frac{\partial^2\sigma}{\partial \epsilon^2} d\epsilon ^2
+ \left( \epsilon \frac{\partial\sigma}{\partial \epsilon} +
\omega \frac{\partial \sigma}{\omega} \right)
\frac{\partial^2 \sigma}{\partial\epsilon\partial \omega} d\epsilon d\omega
+ \omega \frac{\partial \sigma}{\partial \omega}
\frac{\partial^2 \sigma}{\partial\omega^2} d\omega ^2\ .
\ee
Then, for the van der Waals fundamental equation we get
\be
g^{III} = \frac{9}{2(\epsilon\omega+3)^3} \left[ -\epsilon \omega^2 d\epsilon^2 +
\frac{3(5\epsilon \omega^2 -\epsilon\omega-3\omega +3)}{3\omega -1} d\epsilon d\omega
+ \frac{3f(\epsilon,\omega)}{2(3\omega-1)} d\omega^2 \right] \ .
\ee
The corresponding curvature cannot be put in a compact form. However, one can show that in general the curvature singularities of the metric $g^{III}$ are determined by the condition
\be
\frac{\partial ^2 \sigma}{\partial \epsilon \partial \omega} = 0
\ee
which is not satisfied by the van der Waals fundamental equation (\ref{reducedentropy}).

We conclude that in this case all the information about the curvature singularities of the equilibrium space is contained in the metrics $g^I$ and $g^{II}$ only.

\section{Interpretation of the results }
\label{sec:phases}

According to the formalism of GTD, curvature singularities should indicate the presence of phase transitions. This intuitive idea is based upon the fact that singularities are critical locations at which the concepts of differential geometry cannot be applied anymore, i.e., singularities indicate a break down of the theory of differential geometry. On the other hand, we know that equilibrium thermodynamics breaks down when the system undergoes phase transitions. Based on this analogy, the geometric approaches to thermodynamics consider curvature singularities as the geometric representation of phase transitions. We will show this in the case of van der Waals systems.

\subsection{First order phase transitions}

\begin{figure}[h]{\includegraphics[scale=0.35]{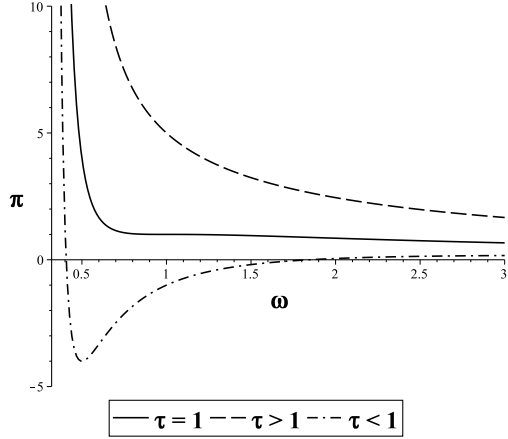}
        \caption{Isotherms by the van der Waals gas in reduced variables
        }
        \label{fig:isoth}
    }
\end{figure}

First order phase transitions can be detected by considering the isotherms of the reduced pressure (\ref{reducepressure}), which are illustrated in the $\pi-\omega$ plots of Fig. \ref{fig:isoth}.

 We can see that the behavior of the isotherms depends on the value of $\tau$ and  $\tau=1$ determines the limit between two different classes of isotherms. This is the isotherm for which the temperature coincides with the critical temperature; in Fig. \ref{fig:isoth}, it corresponds to the curve with an inflection point, i.e.,
\bea \label{criticalpoints} \Bigg(\frac{d\pi}{d\omega}\Bigg)_\tau
= \Bigg(\frac{d^2 \pi}{d\omega^2}\Bigg)_\tau=0\,,\eea
Then, using the equation of state  given in (\ref{stateequation}), one can show that the inflection point is determined by the solutions of the algebraic equation
\be \label{criticalpoints2}
\epsilon \omega^3-6\omega^2+6\omega-1=0\,,\ee
which are also interpreted as the points where first order phase transitions occur \cite{callen}. This is exactly the condition we have found in Eq.(\ref{curvsingI}) for the existence of curvature singularities of the metric $g^I$.

\subsection{Second order phase transitions}

The response functions of a system are used to indicate the
presence of second order phase transitions. In the case of a
simplet system with two thermodynamic degrees of freedom, there
exist only three independent response functions \cite{callen}; for
concreteness,  we choose as independent  functions the isothermal
compressibility $\alpha $, the thermal expansion at constant
pressure $\beta$, and the heat capacity at constant pressure and
temperature $c$, which are  defined \be \alpha=\frac{1}{v}\left(
\frac{\partial
    v}{\partial T}\right)_p\,,\ \
 \beta=-\frac{1}{v}\left(
\frac{\partial v}{\partial p}\right)_T\, , \ \ c=
T\frac{\left(\frac{\partial v}{\partial
T}\right)^2_p}{\left(\frac{\partial v}{\partial p}\right)_T} \ .
\ee Using the relationships: $v=v_c \omega$, $T=T_c \tau$, and
$p=p_c\pi$, we obtain the reduced response functions \be
\label{reducedalpha} \tilde{\alpha}=\frac{1}{\omega}\left(
\frac{\partial \omega}{\partial \tau}\right)_\pi\,,\ \
 \tilde{\beta}=-\frac{1}{\omega}\left(
\frac{\partial \omega}{\partial \pi}\right)_\tau\, , \ \ \tilde c=
-\frac{3\tau}{8}\frac{\left(\frac{\partial \omega}{\partial
\tau}\right)^2_\pi}{\left(\frac{\partial \omega}{\partial
\pi}\right)_\tau} \ . \ee

Furthermore, the reduced response functions can be expressed in
terms of the derivatives of the fundamental equation as \be
\tilde{\alpha} =-\frac{8}{3\pi}\frac{\left(
    \frac{\partial \sigma}{\partial \omega}\right)^2_\pi}{\left(
    \frac{\partial^2 \sigma}{\partial \omega^2}\right)_\pi} \,,
 \ \
 \tilde{\beta}
=-\frac{3}{8\omega \tau}\frac{1}{\left( \frac{\partial^2
        \sigma}{\partial \omega^2}\right)_\tau} \,, \ \
\tilde c = -\frac{64 \tau^2 \Big(\frac{\partial \sigma}{\partial
\omega}\Big)^{4}_\epsilon }{9 \pi^2 \Big(\frac{\partial^2
\sigma}{\partial \omega^2}\Big)_\epsilon}\,. \ee Then, substituing
the corresponding quantities for a van der Waals system, we obtain
\bea \label{alpha4} \tilde{\alpha} &=&\frac{4 \omega
    (3\omega-1)(2\epsilon \omega^2-3\omega+3)}{3(2\epsilon^2 \omega^4
    -6\epsilon \omega^3 +12 \epsilon \omega^2-2\epsilon \omega
    -9\omega^2+18\omega-3)}\,,
\\\label{beta4}\tilde{\beta} &=&\frac{ \omega^2
    (3\omega-1)^2(\epsilon \omega+3)}{3(2\epsilon^2 \omega^4
    -6\epsilon \omega^3 +12 \epsilon \omega^2-2\epsilon \omega
    -9\omega^2+18\omega-3)} \,,\\
\tilde c &=& \frac{
 (2\epsilon \omega^2-3\omega+3)^2}{2[2\epsilon^2 \omega^4-6\epsilon \omega^3+12 \epsilon
\omega^2-2\epsilon\omega-9\omega^2+18\omega-3]}\ .
 \eea
It follows from the above expressions that the response functions
diverge if the condition \be 2\epsilon^2 \omega^4 -6\epsilon
\omega^3 +12 \epsilon \omega^2-2\epsilon \omega
-9\omega^2+18\omega-3 = 0 \label{cond2} \ee is satisfied. We
recognize immediately that this condition is identical to the
condition for the existence of curvature singularities in the
equilibrium space of the metric $g^{II}$, as given in
Eq.(\ref{singII}).

\section{Thermodynamic geometry of van der Waals systems}
\label{sec:tdg}

Consider a system described by the fundamental equation $\Phi=\Phi(E^a)$.
Thermodynamic geometry is an approach based upon the assumption that the geometric properties of the corresponding equilibrium space are determined by a Hessian metric
\be
g_{ab}^H = \frac{\partial \Phi^2}{\partial E^a \partial E^b}\ .
\ee
Although a Hessian metric can be constructed for any function $\Phi$, in thermodynamics only two options have been studied. If the thermodynamic potential is chosen as minus the entropy, $-S$,  or the internal energy, $U$, the resulting Hessian metrics
\be
g_{ab}^R = - \frac{\partial S^2}{\partial E^a \partial E^b}\ , \ \
g_{ab}^W = \frac{\partial U^2}{\partial E^a \partial E^b}\ ,
\ee
are called Ruppeiner \cite{rup79} and Weinhold \cite{wei75} metrics. We will now investigate the curvature of these metrics in the case of van der Waals systems by using reduced variables.

Using the fundamental equation (\ref{reducedentropy}), we obtain
\be
g^R = \frac{3}{2(\epsilon\omega+3)}\left[ -\frac{\omega^2}{\epsilon\omega+3} d\epsilon^2 + 6 d\epsilon d\omega - \frac{9}{2} h(\epsilon,\omega) d\omega^2\right] \,
\ee
where
\be
h(\epsilon,\omega)=  \frac{ 2\epsilon^2 \omega^4 - 6\epsilon \omega^3 + 12 \epsilon \omega^2 - 2\epsilon \omega - 9\omega^2 + 18 \omega -3}{ \omega^2(\epsilon\omega+3)(3\omega-1)^2} \ ,
\ee
and the corresponding curvature scalar reads
\be
R^R = - \frac{(3\omega-1)^4 \omega^2 h(\epsilon,\omega)}{3 (\epsilon \omega^3 - 6 \omega^2+ 6\omega -1)} \ .
\ee

Consider now the Weinhold metric. In the energy representation, the fundamental equation (\ref{reducedentropy}) can be expressed as
\be
\epsilon = \frac{ e^{2/3(\sigma-\sigma_0)} }{(\omega - 1/3)^{2/3}} - \frac{3}{\omega} \ .
\ee
Then, the Weinhold metric can be written explicitly written as
\bea
g^W &=& \frac{4}{9} \left(\epsilon + 3/\omega\right) d\sigma^2
- \frac{8}{3} \frac{ \epsilon +3/\omega }{3\omega -1} d\sigma d\omega \nonumber \\
&+& \frac{2(5\epsilon\omega^2 - 12\omega^2+18\omega -3)}{\omega^3(3\omega-1)^2} d\omega^2\ ,
\eea
from which we compute the curvature scalar and get
\be
R^W = - \frac{3\omega^3 (3\omega-1)^2 }{6(\epsilon\omega^3 - 6\omega^2 + 6\omega -1)}\ .
\ee

From the above expressions for the curvature scalar it follows that the condition
\be
\epsilon\omega^3 - 6\omega^2 + 6\omega -1 = 0
\ee
indicates the presence of singularities in the equilibrium space described by the Ruppeiner and Weinhold metrics. According to the results presented in Sec. \ref{sec:phases}, the curvature of the Hessian metrics can reproduce the first order phase transitions determined by the condition (\ref{criticalpoints2}), but they are not able to reproduce the second order phase transitions that are predicted by the response functions in Eq.(\ref{cond2}).


\section{Conclusions}
\label{sec:con}

In this work, we tested the formalism of GTD in the case of van der Waals systems. To this end, we first derived the reduced form of the fundamental equation, which has the advantage of being completely general in the sense that it does not depend on the particular values of the van der Waals constants $a$ and $b$.

We calculated the scalar curvature of all the three GTD metrics in order to find the locations where curvature singularities can exist. This analysis lead to the conclusion that there are two different conditions that indicate the presence of singularities. By analyzing the thermodynamic properties of van der Waals systems we proved that the locations of the curvature singularities coincide with the points at which first and second order phase transitions occur. This shows that GTD can describe the entire phase transition structure of van der Waals systems.

Moreover, we investigated the properties of the equilibrium space as defined in thermodynamic geometry by means of Hessian metrics. We found that this approach correctly predicts the presence of first order phase transitions only.

The results presented in this work can be applied to any system characterized by two thermodynamic degrees of freedom. The formalism of GTD is completely general, although the explicit form of the GTD metrics depend on the fundamental equation of the corresponding system.
We thus conclude that to study  the equilibrium space and the complete phase transition structure of any thermodynamic system in an invariant way, one can use the metrics obtained in the framework of the GTD formalism.

\section*{Acknowledgments}

This work was carried out within the scope of the project CIAS 3131
supported by the Vicerrector\'\i a de Investigaciones de la Universidad
Militar Nueva Granada - Vigencia 2020.
This work was partially supported  by UNAM-DGAPA-PAPIIT, Grant No. 114520, Conacyt-Mexico, Grant No. A1-S-31269,
 and by the Ministry of Education and Science of RK, Grant No.
BR05236322 and AP05133630.


\begin{thebibliography}{21}%
	\makeatletter
	\providecommand \@ifxundefined [1]{%
		\@ifx{#1\undefined}
	}%
	\providecommand \@ifnum [1]{%
		\ifnum #1\expandafter \@firstoftwo
		\else \expandafter \@secondoftwo
		\fi
	}%
	\providecommand \@ifx [1]{%
		\ifx #1\expandafter \@firstoftwo
		\else \expandafter \@secondoftwo
		\fi
	}%
	\providecommand \natexlab [1]{#1}%
	\providecommand \enquote  [1]{``#1''}%
	\providecommand \bibnamefont  [1]{#1}%
	\providecommand \bibfnamefont [1]{#1}%
	\providecommand \citenamefont [1]{#1}%
	\providecommand \href@noop [0]{\@secondoftwo}%
	\providecommand \href [0]{\begingroup \@sanitize@url \@href}%
	\providecommand \@href[1]{\@@startlink{#1}\@@href}%
	\providecommand \@@href[1]{\endgroup#1\@@endlink}%
	\providecommand \@sanitize@url [0]{\catcode `\\12\catcode `\$12\catcode
		`\&12\catcode `\#12\catcode `\^12\catcode `\_12\catcode `\%12\relax}%
	\providecommand \@@startlink[1]{}%
	\providecommand \@@endlink[0]{}%
	\providecommand \url  [0]{\begingroup\@sanitize@url \@url }%
	\providecommand \@url [1]{\endgroup\@href {#1}{\urlprefix }}%
	\providecommand \urlprefix  [0]{URL }%
	\providecommand \Eprint [0]{\href }%
	\providecommand \doibase [0]{http://dx.doi.org/}%
	\providecommand \selectlanguage [0]{\@gobble}%
	\providecommand \bibinfo  [0]{\@secondoftwo}%
	\providecommand \bibfield  [0]{\@secondoftwo}%
	\providecommand \translation [1]{[#1]}%
	\providecommand \BibitemOpen [0]{}%
	\providecommand \bibitemStop [0]{}%
	\providecommand \bibitemNoStop [0]{.\EOS\space}%
	\providecommand \EOS [0]{\spacefactor3000\relax}%
	\providecommand \BibitemShut  [1]{\csname bibitem#1\endcsname}%
	\let\auto@bib@innerbib\@empty
	\bibitem [{\citenamefont {Frankel}(2004)}]{frankel}%
	\BibitemOpen
	\bibfield  {author} {\bibinfo {author} {\bibfnamefont {T.}~\bibnamefont
			{Frankel}},\ }\href {https://books.google.com.mx/books?id=DUnjs6nEn8wC}
	{\emph {\bibinfo {title} {The Geometry of Physics: An Introduction}}}\
	(\bibinfo  {publisher} {Cambridge University Press},\ \bibinfo {year}
	{2004})\BibitemShut {NoStop}%
	\bibitem [{\citenamefont {Rao}(1945)}]{rao45}%
	\BibitemOpen
	\bibfield  {author} {\bibinfo {author} {\bibfnamefont {C.~R.}\ \bibnamefont
			{Rao}},\ }\href@noop {} {\bibfield  {journal} {\bibinfo  {journal} {Bulletin
				of Calcutta Mathematical Society}\ }\textbf {\bibinfo {volume} {37}},\
		\bibinfo {pages} {81} (\bibinfo {year} {1945})}\BibitemShut {NoStop}%
	\bibitem [{\citenamefont {Amari}(2012)}]{amari2012}%
	\BibitemOpen
	\bibfield  {author} {\bibinfo {author} {\bibfnamefont {S.}~\bibnamefont
			{Amari}},\ }\href {https://books.google.com.mx/books?id=XiDnBwAAQBAJ} {\emph
		{\bibinfo {title} {Differential-Geometrical Methods in Statistics}}},\
	Lecture Notes in Statistics\ (\bibinfo  {publisher} {Springer New York},\
	\bibinfo {year} {2012})\BibitemShut {NoStop}%
	\bibitem [{\citenamefont {Weinhold}(1975)}]{wei75}%
	\BibitemOpen
	\bibfield  {author} {\bibinfo {author} {\bibfnamefont {F.}~\bibnamefont
			{Weinhold}},\ }\href@noop {} {\bibfield  {journal} {\bibinfo  {journal} {The
				Journal of Chemical Physics}\ }\textbf {\bibinfo {volume} {63}},\ \bibinfo
		{pages} {2479} (\bibinfo {year} {1975})}\BibitemShut {NoStop}%
	\bibitem [{\citenamefont {Ruppeiner}(1979)}]{rup79}%
	\BibitemOpen
	\bibfield  {author} {\bibinfo {author} {\bibfnamefont {G.}~\bibnamefont
			{Ruppeiner}},\ }\href@noop {} {\bibfield  {journal} {\bibinfo  {journal}
			{Physical Review A}\ }\textbf {\bibinfo {volume} {20}},\ \bibinfo {pages}
		{1608} (\bibinfo {year} {1979})}\BibitemShut {NoStop}%
	\bibitem [{\citenamefont {Ruppeiner}(1995)}]{rup95}%
	\BibitemOpen
	\bibfield  {author} {\bibinfo {author} {\bibfnamefont {G.}~\bibnamefont
			{Ruppeiner}},\ }\href@noop {} {\bibfield  {journal} {\bibinfo  {journal}
			{Reviews of Modern Physics}\ }\textbf {\bibinfo {volume} {67}},\ \bibinfo
		{pages} {605} (\bibinfo {year} {1995})}\BibitemShut {NoStop}%
	\bibitem [{\citenamefont {Quevedo}(2007)}]{quev07}%
	\BibitemOpen
	\bibfield  {author} {\bibinfo {author} {\bibfnamefont {H.}~\bibnamefont
			{Quevedo}},\ }\href@noop {} {\bibfield  {journal} {\bibinfo  {journal}
			{Journal of Mathematical Physics}\ }\textbf {\bibinfo {volume} {48}},\
		\bibinfo {pages} {013506} (\bibinfo {year} {2007})}\BibitemShut {NoStop}%
	\bibitem [{\citenamefont {Callen}(1985)}]{callen}%
	\BibitemOpen
	\bibfield  {author} {\bibinfo {author} {\bibfnamefont {H.~B.}\ \bibnamefont
			{Callen}},\ }\href {https://cds.cern.ch/record/450289} {\emph {\bibinfo
			{title} {{Thermodynamics and an introduction to thermostatistics; 2nd
					ed.}}}}\ (\bibinfo  {publisher} {Wiley},\ \bibinfo {address} {New York, NY},\
	\bibinfo {year} {1985})\BibitemShut {NoStop}%
	\bibitem [{\citenamefont {Weinhold}(2009)}]{weinholdbook}%
	\BibitemOpen
	\bibfield  {author} {\bibinfo {author} {\bibfnamefont {F.}~\bibnamefont
			{Weinhold}},\ }\href@noop {} {\emph {\bibinfo {title} {Classical and
				geometrical theory of chemical and phase thermodynamics}}}\ (\bibinfo
	{publisher} {John Wiley \& Sons},\ \bibinfo {year} {2009})\BibitemShut
	{NoStop}%
	\bibitem [{\citenamefont {Ruppeiner}(2014)}]{rup14}%
	\BibitemOpen
	\bibfield  {author} {\bibinfo {author} {\bibfnamefont {G.}~\bibnamefont
			{Ruppeiner}},\ }in\ \href@noop {} {\emph {\bibinfo {booktitle} {Breaking of
				Supersymmetry and Ultraviolet Divergences in Extended Supergravity}}}\
	(\bibinfo  {publisher} {Springer},\ \bibinfo {year} {2014})\ pp.\ \bibinfo
	{pages} {179--203}\BibitemShut {NoStop}%
	\bibitem [{\citenamefont {{\AA}man}\ \emph {et~al.}(2015)\citenamefont
		{{\AA}man}, \citenamefont {Bengtsson},\ and\ \citenamefont
		{Pidokrajt}}]{abp15}%
	\BibitemOpen
	\bibfield  {author} {\bibinfo {author} {\bibfnamefont {J.~E.}\ \bibnamefont
			{{\AA}man}}, \bibinfo {author} {\bibfnamefont {I.}~\bibnamefont {Bengtsson}},
		\ and\ \bibinfo {author} {\bibfnamefont {N.}~\bibnamefont {Pidokrajt}},\
	}\href@noop {} {\bibfield  {journal} {\bibinfo  {journal} {Entropy}\ }\textbf
		{\bibinfo {volume} {17}},\ \bibinfo {pages} {6503} (\bibinfo {year}
		{2015})}\BibitemShut {NoStop}%
	\bibitem [{\citenamefont {Mandal}\ and\ \citenamefont
		{Biswas}(2015)}]{manbis15}%
	\BibitemOpen
	\bibfield  {author} {\bibinfo {author} {\bibfnamefont {A.}~\bibnamefont
			{Mandal}}\ and\ \bibinfo {author} {\bibfnamefont {R.}~\bibnamefont
			{Biswas}},\ }\href@noop {} {\bibfield  {journal} {\bibinfo  {journal}
			{Astrophysics and Space Science}\ }\textbf {\bibinfo {volume} {357}},\
		\bibinfo {pages} {1} (\bibinfo {year} {2015})}\BibitemShut {NoStop}%
	\bibitem [{\citenamefont {Bravetti}\ and\ \citenamefont
		{Luongo}(2014)}]{braluo14}%
	\BibitemOpen
	\bibfield  {author} {\bibinfo {author} {\bibfnamefont {A.}~\bibnamefont
			{Bravetti}}\ and\ \bibinfo {author} {\bibfnamefont {O.}~\bibnamefont
			{Luongo}},\ }\href@noop {} {\bibfield  {journal} {\bibinfo  {journal}
			{International Journal of Geometric Methods in Modern Physics}\ }\textbf
		{\bibinfo {volume} {11}},\ \bibinfo {pages} {1450071} (\bibinfo {year}
		{2014})}\BibitemShut {NoStop}%
	\bibitem [{\citenamefont {Hendi}\ \emph {et~al.}(2016)\citenamefont {Hendi},
		\citenamefont {Panahiyan},\ and\ \citenamefont {Panah}}]{hendi16}%
	\BibitemOpen
	\bibfield  {author} {\bibinfo {author} {\bibfnamefont {S.}~\bibnamefont
			{Hendi}}, \bibinfo {author} {\bibfnamefont {S.}~\bibnamefont {Panahiyan}}, \
		and\ \bibinfo {author} {\bibfnamefont {B.~E.}\ \bibnamefont {Panah}},\
	}\href@noop {} {\bibfield  {journal} {\bibinfo  {journal} {Journal of High
				Energy Physics}\ }\textbf {\bibinfo {volume} {2016}},\ \bibinfo {pages} {1}
		(\bibinfo {year} {2016})}\BibitemShut {NoStop}%
	\bibitem [{\citenamefont {Kubiz{\v{n}}{\'a}k}\ and\ \citenamefont
		{Mann}(2015)}]{kubman15}%
	\BibitemOpen
	\bibfield  {author} {\bibinfo {author} {\bibfnamefont {D.}~\bibnamefont
			{Kubiz{\v{n}}{\'a}k}}\ and\ \bibinfo {author} {\bibfnamefont {R.~B.}\
			\bibnamefont {Mann}},\ }\href@noop {} {\bibfield  {journal} {\bibinfo
			{journal} {Canadian Journal of Physics}\ }\textbf {\bibinfo {volume} {93}},\
		\bibinfo {pages} {999} (\bibinfo {year} {2015})}\BibitemShut {NoStop}%
	\bibitem [{\citenamefont {Arnold}(1989)}]{arnold}%
	\BibitemOpen
	\bibfield  {author} {\bibinfo {author} {\bibfnamefont {V.}~\bibnamefont
			{Arnold}},\ }\href@noop {} {\emph {\bibinfo {title} {Mathematical methods of
				classical mechanics}}},\ Vol.~\bibinfo {volume} {60}\ (\bibinfo  {publisher}
	{Springer},\ \bibinfo {year} {1989})\BibitemShut {NoStop}%
	\bibitem [{\citenamefont {Alberty}(1994)}]{alberty1994}%
	\BibitemOpen
	\bibfield  {author} {\bibinfo {author} {\bibfnamefont {R.~A.}\ \bibnamefont
			{Alberty}},\ }\href {\doibase 10.1021/cr00030a001} {\bibfield  {journal}
		{\bibinfo  {journal} {Chemical Reviews}\ }\textbf {\bibinfo {volume} {94}},\
		\bibinfo {pages} {1457} (\bibinfo {year} {1994})},\ \Eprint
	{http://arxiv.org/abs/https://doi.org/10.1021/cr00030a001}
	{https://doi.org/10.1021/cr00030a001} \BibitemShut {NoStop}%
	\bibitem [{\citenamefont {Quevedo}\ \emph {et~al.}(2018)\citenamefont
		{Quevedo}, \citenamefont {Quevedo},\ and\ \citenamefont
		{S{\'a}nchez}}]{qqs19}%
	\BibitemOpen
	\bibfield  {author} {\bibinfo {author} {\bibfnamefont {H.}~\bibnamefont
			{Quevedo}}, \bibinfo {author} {\bibfnamefont {M.~N.}\ \bibnamefont
			{Quevedo}}, \ and\ \bibinfo {author} {\bibfnamefont {A.}~\bibnamefont
			{S{\'a}nchez}},\ }\href@noop {} {\bibfield  {journal} {\bibinfo  {journal}
			{The European Physical Journal C}\ }\textbf {\bibinfo {volume} {79}},\
		\bibinfo {pages} {1} (\bibinfo {year} {2018})}\BibitemShut {NoStop}%
	\bibitem [{\citenamefont {Greiner}\ \emph {et~al.}(2012)\citenamefont
		{Greiner}, \citenamefont {Neise},\ and\ \citenamefont
		{St{\"o}cker}}]{greiner}%
	\BibitemOpen
	\bibfield  {author} {\bibinfo {author} {\bibfnamefont {W.}~\bibnamefont
			{Greiner}}, \bibinfo {author} {\bibfnamefont {L.}~\bibnamefont {Neise}}, \
		and\ \bibinfo {author} {\bibfnamefont {H.}~\bibnamefont {St{\"o}cker}},\
	}\href@noop {} {\emph {\bibinfo {title} {Thermodynamics and statistical
				mechanics}}}\ (\bibinfo  {publisher} {Springer Science \& Business Media},\
	\bibinfo {year} {2012})\BibitemShut {NoStop}%
	\bibitem [{\citenamefont {Huang}(2009)}]{huang}%
	\BibitemOpen
	\bibfield  {author} {\bibinfo {author} {\bibfnamefont {K.}~\bibnamefont
			{Huang}},\ }\href@noop {} {\emph {\bibinfo {title} {Introduction to
				statistical physics}}}\ (\bibinfo  {publisher} {Chapman and Hall/CRC},\
	\bibinfo {year} {2009})\BibitemShut {NoStop}%
	\bibitem [{\citenamefont {Reif}(2009)}]{reif}%
	\BibitemOpen
	\bibfield  {author} {\bibinfo {author} {\bibfnamefont {F.}~\bibnamefont
			{Reif}},\ }\href@noop {} {\emph {\bibinfo {title} {Fundamentals of
				statistical and thermal physics}}}\ (\bibinfo  {publisher} {Waveland Press},\
	\bibinfo {year} {2009})\BibitemShut {NoStop}%
\end{thebibliography}


%

\end{document}